\documentclass[twocolumn,hf]{ceurart}
\usepackage{scrlayer-scrpage}
\clearpairofpagestyles % Clear default header and footer

\ihead{Louise McCormack and Malika Bendechache} % Left header
\ofoot[\pagemark]{\pagemark} % Centered footer with page number

\usepackage{graphicx} % For including graphics
\usepackage{amsmath}  % For advanced math typesetting
\usepackage{natbib}   % For bibliography management

\usepackage{listings}
\begin{document}

\copyrightyear{2024}
\copyrightclause{Copyright for this paper by its authors.
  Use permitted under Creative Commons License Attribution 4.0
  International (CC BY 4.0).}

\conference{Accepted for presentation at AIEB 2024: Workshop on Implementing AI Ethics Through a Behavioural Lens - ECAI 2024}

\title{Ethical AI Governance: Methods for Evaluating Trustworthy AI}

\author[1]{Louise McCormack}[%
orcid=0009-0001-1782-1214,
email=louise.mccormack@adaptcentre.ie,
]
\address[1]{ADAPT Research Centre, University of Galway, Ireland}

\author[1]{Malika Bendechache}[%
orcid=0000-0003-0069-1860,
email=malika.bendechache@adaptcentre.ie,
]
\fnmark[1]

\cortext[1]{Corresponding author.}
\fntext[1]{These authors contributed equally.}

\begin{abstract}
  Trustworthy Artificial Intelligence (TAI) integrates ethics that align with human values, looking at their influence on AI behaviour and decision-making. Primarily dependent on self-assessment, TAI evaluation aims to ensure ethical standards and safety in AI development and usage. This paper reviews the current TAI evaluation methods in the literature and offers a classification, contributing to understanding self-assessment methods in this field.
\end{abstract}


%%
%% This command processes the author affiliation and title
\maketitle
\section{Introduction}

Artificial intelligence (AI) is increasingly integrated into numerous sectors, making ethical considerations and trustworthiness in AI systems more critical than ever. Behavioural science is utilised to achieve objectives in areas such as climate change mitigation and educational attainment\cite{hallsworth2023}, a trend which also extends to Trustworthy AI (TAI). TAI is a crucial concept within the field of ethical AI, which encompasses the ethical considerations essential in the development and use of AI systems\cite{Kauretal_2022}. Leading TAI frameworks\cite{Floridietal_2018}\cite{thiebes2021}\cite{kaur2021} incorporate behavioural science principles to ensure AI systems align with human values, considering their impact on behaviour and decision-making. Additionally, bidirectional human-AI alignment emphasises aligning AI to human values and enabling humans to adjust to AI advancements cognitively and behaviourally\cite{shen2024}.

The European Commission Assessment List for Trustworthy AI (ALTAI)\cite{EUALTAI_2020} and the European Union (EU) AI Act\cite{eu2024aiact} are essential TAI guidelines, emphasising a human-centred, interdisciplinary approach. One recommended governance approach is establishing Standard-Setting Organisations that ensure minimum standards for testing, documentation and public reporting\cite{laux2023three}. Despite the availability of various standards such as ISO/IEC 42001\cite{ISO42001}, evaluating and auditing AI systems remains challenging. 

Several key surveys, such as those by Liu et al.\cite{Liuetal_2022} and Chamolaetal et al.\cite{Chamolaetal_2023}, compile summaries of existing technical methods and technology in TAI. However, these surveys do not focus on methods to score the areas of TAI. Ojewale et al.\cite{ojewale2024towards} propose a process for AI auditing, and although this work highlights the need for metrics and standards, it does not delve into the methods for calculating such metrics.

In this paper, we summarise and propose a classification and sub-classification for existing methods and systems to govern, evaluate, and score AI systems for trustworthiness aligned with the interdisciplinary human-centred approach taken by the EU. We also discuss challenges and future work in this area.

%%%%%%%%%%%%%%%%%%%%%%%%%%%%%%%%%%%%%%%%%%%%%%%%%%%%%%%%%%%%%%%%%%%%%%%%

\section{Methodology}
\subsection{Review Technique}
Our survey was conducted through a Google Scholar query to identify methods used in the literature for TAI evaluation. In addition, we added articles, regulatory documentation, and ISO standards in this area through snowballing.

\subsection{Research Questions}
The following are the identified research questions for this review:
\begin{itemize}
    \item Q1: What TAI evaluation methods and systems exist in the literature?
    \item Q2: What barriers to evaluating TAI are highlighted in the literature? 
\end{itemize}

\subsection{Research Search and Data Extraction Strategy}
A search string for Google Scholar was designed to capture papers discussing topics in machine learning, trust and evaluation areas. Two researchers independently screened titles first and the abstract second to find papers that included TAI evaluation methods, resulting in 380 papers from the search string and an additional 12 papers through snowballing being reviewed. These papers were narrowed further, bringing the number of papers contributing to the core findings to 34. These papers were then summarised by both researchers and used to create a classification for the TAI evaluation methods.

\section{Methods for Evaluating Trustworthy AI}
In this section, we propose a classification for evaluating and scoring TAI. Of the papers reviewed, we found several approaches to AI scoring methods that considered various areas within TAI. Based on maturity and the type of solution proposed, we classed these papers into four categories: conceptual evaluation methods, Manual evaluation methods, Automated Evaluation Methods and Semi-Automated Evaluation Methods. In addition to this, we proposed a sub-classification based on the topic being evaluated. These sub-classifications are fairness \& compliance evaluation, transparency evaluation, risk \& accountability evaluation and trust \& safety evaluation. As outlined in Figure 1, the most common approaches are conceptual approaches, indicating the lack of maturity in this field. This figure also shows the breakdown of evaluation approaches by topic, particularly the number of automated and semi-automated evaluation methods already developed in fairness and compliance, one of the more researched areas of trustworthy AI.

\begin{figure*}
  \centering
  \includegraphics[width=\linewidth]{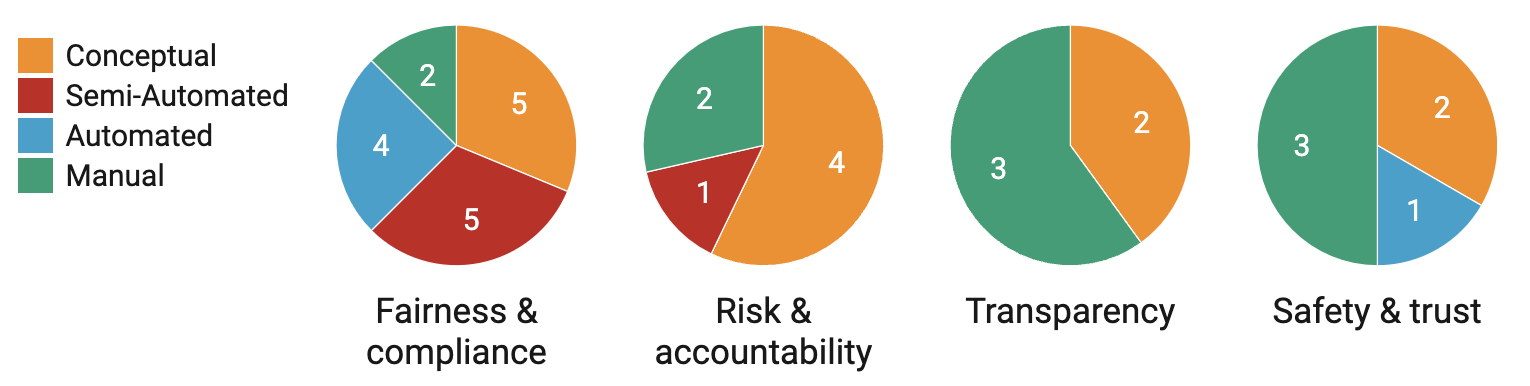}
  \caption{Comparison of evaluation methods by topic}
\end{figure*}

\subsection{Conceptual Evaluation Methods}
The existing research includes several high-level governance frameworks that consider multiple dimensions of trustworthy AI throughout the AI lifecycle. Conceptual evaluation methods are high-level methods that do not provide implementation details or are not tested and validated. While conceptual frameworks in the literature can be holistic, they can also lack detail.

\subsubsection{Fairness \& Compliance Evaluation}
Several conceptual approaches sought to evaluate and improve fairness and compliance in AI systems, introducing concepts like policy violation detection\cite{Shaikhetal_2017}, using AI to define ethical behaviour\cite{buenfiletal_2019}\cite{UmbrelloVanDePoel2021} and automating fairness auditing\cite{park2022fairness}\cite{van2022algorithmic}. Researchers used a variety of approaches in deciding what was fair, including incorporating existing established ethical guidelines\cite{UmbrelloVanDePoel2021}, extracting ethical guidelines from social media\cite{buenfiletal_2019}, using a third-party regulator\cite{van2022algorithmic},\cite{park2022fairness} and extracting guidelines from policy documents\cite{Shaikhetal_2017}. 

\subsubsection{Transparency Evaluation}
Researchers proposed approaches that included evaluating transparency in areas such as healthcare\cite{Jiaetal_2022} and finance\cite{Lee_2019}.
The proposed framework by Lee\cite{Lee_2019} involved scoring fairness and interoperability, allowing humans to oversee and make conscious choices affecting both. The approach is context-conscious fairness and considers the trade-off between accuracy and interpretability and the trade-off between aggregate benefit and inequity. Trade-offs are benchmarked to make transparent, context-based, informed choices when using Machine Learning (ML) for decision-making. Jia et al.\cite{Jiaetal_2022} proposed a framework to measure and improve technical robustness, safety, and transparency. It involved quantifying performance and XAI and establishing a trade-off between these trust properties for the ML algorithm selection for their healthcare use case. 

\subsubsection{Risk \& Accountability Evaluation}
Researchers also proposed conceptual governance frameworks that focused on risk management and accountability. These included ethical AI risk evaluation frameworks that built on the existing concepts such as operational design domain (ODD)\cite{roski2021enhancing}\cite{mattioli2024}. The importance of defined safety boundaries was also highlighted\cite{stettinger2024}\cite{mattioli2024}. 

Lu et al.\cite{lu2024responsible} published a Responsible Artificial Intelligence (RAI) Pattern Catalogue, which was divided into multi-level governance patterns, trustworthy process patterns, and RAI-by-design product patterns, considering stakeholders at the industry, organisation, and team levels. This is important as researchers have shown engineers, legal experts, and users all require different levels of transparency from AI systems\cite{van2020transparency}.

\subsubsection{Trust \& Safety Evaluation}
Conceptual evaluation frameworks also addressed trust\cite{Umetal_2022}\cite{broderick2023toward} and safety\cite{Fisher2020}. These frameworks firstly focused on identifying evaluation criteria or trust risk areas, then on methods to address these risk areas to improve trust\cite{Umetal_2022}\cite{broderick2023toward}. 

Fisher et al.\cite{Fisher2020} discuss several use cases, focusing on safety-critical domains that require new standards and verification, validation, and certification methods. They include a classification of verification methods, including formal exhaustive static methods like model checking and theorem proving, non-exhaustive dynamic semi-formal methods like runtime verification and software testing, and non-exhaustive static methods like static analysis. The paper highlights the difficulty in certifying autonomous systems due to their complexity and evolving nature. Multiple stakeholder involvement creates complexity in establishing a consensus on acceptable ethical standards or evaluation criteria that do not disclose sensitive information.

Um et al.'s\cite{Umetal_2022} layered trust framework includes a Trust Agent for data extraction, a Trust Analysis layer for computing trust metrics, and a Trust Management layer, addressing risk, fairness, security, design, traceability, data security, data privacy, and data pre-processing. Broderick et al.\cite{broderick2023toward}
created a taxonomy of trust in AI, which includes a process diagram for assessing the areas in which trust in ML can fail. They considered real-world use cases for finance, healthcare, and politics and subsequently provided ways to mitigate the risk and increase trust at each stage. Their conceptual process seeks to assess and mitigate the level of user trust, specifically the trust of an expert in their field at each stage. 

\subsection{Manual Evaluation Methods}
One method proposed for assessing TAI is a manual questionnaire. Beyond the questions from the EU ALTAI\cite{EUALTAI_2020} and ISO/IEC standards\cite{iso27001}\cite{ISO42001}, six additional questionnaires were identified to score AI systems for trustworthiness. Manual questionnaires align with this area's regulation, considering multiple EU TAI principles. The disadvantage of the manual approach is that these questionnaires are typically time-consuming. This can lead to business constraints in completing the questions due to limited information about the external data the systems used\cite{Fehretal_2022}.

\subsubsection{Fairness \& Compliance Evaluation} 
Approaches to improve fairness and achieve compliance in machine learning were proposed by researchers\cite{Leeetal_2021}\cite{LandersandBehrend_2022}. One approach was a practical questionnaire to help improve fairness by detecting bias\cite{Leeetal_2021}. A second approach to audit and score fairness in ML considered twelve metrics in this area\cite{LandersandBehrend_2022}. The first six metrics focus on the stages of data collection, model development, feature selection and model performance—three metrics related to the human relationship with the model's decisions or predictions. The final metrics focus on assessing fairness from a broader social impact and include the three meta-components: cultural context, respect, and the research design process.  

\subsubsection{Transparency Evaluation}
Transparency-focused questionnaires that focused on assessing the transparency of several TAI principles were also proposed by some researchers\cite{Chaudhryetal_2022}\cite{Fehretal_2022}\cite{Bommasanietal_2023}. A notable questionnaire in the area of transparency is Bommasani et al.\cite{Bommasanietal_2023} who proposed The Foundation Model Transparency Index (FMTI), which included 100 indicators for transparency to be self-scored using a three-tier questionnaire and included benchmarks for leading organisations such as Open AI, AWS and Meta. Other researchers created separate transparency criteria for different tiers of stakeholders\cite{Chaudhryetal_2022} and proposed using weighted questions using a 3-point scale for each question\cite{Fehretal_2022}. Transparency was also a consideration by researchers who looked at other areas such as user trust\cite{Guoetal_2022}. 

\subsubsection{Risk \& Accountability Evaluation}
For security evaluation, researchers\cite{mcintosh2024} scored existing questionnaire-based frameworks used in industry NIST\cite{nist_framework}, COBIT\cite{cobit}, ISO27001\cite{iso27001}, and ISO42001\cite{ISO42001} for their potential usage for AI's that incorporate Large Language Models (LLMs). Additionally, researchers developed a framework to evaluate the MITRE ATLAS\cite{mitreatlas} framework's effectiveness in protecting ML systems from poisoning attacks, scoring multiple TAI principles using a qualitative severity rating scale\cite{wymberry2024approach}.

\subsubsection{Trust \& Safety Evaluation}
Several questionnaire-based papers focused on trust and safety evaluation, typically asking users about their trust in various AI systems\cite{Dvoraketal_2021}\cite{Druceetal_2021}\cite{Guoetal_2022}.

One approach was a simple unweighted user survey-based questionnaire, which scored several aspects of TAI evaluation, including intent and limitations, data, explainability, safety and robustness, audibility, and accountability\cite{Dvoraketal_2021}. Researchers also developed frameworks that used surveys to quantify and improve user trust by improving the transparency of the system\cite{Druceetal_2021}\cite{Guoetal_2022}. Both papers successfully indicated a correlation between increased transparency and increased user trust in AI.

\subsection{Automatic Evaluation Methods}
This section includes papers investigating automated scoring methods for TAI Principles. Automatic methods ensure consistency in evaluation, however they rely on predefined metrics which do not exist for many aspects of trustworthy AI. The automated methods published to date are technical methods to evaluate and score the technical aspects of trustworthy AI with established methods and metrics.

\subsubsection{Fairness \& Compliance Evaluation}
Several automated methods published to date are technical methods to evaluate and score fairness\cite{Khalilietal_2021}\cite{Singhetal_2021}\cite{Barza_2023}. Notable methods include using data sampling techniques to measure and understand root causes of bias\cite{Singhetal_2021} and a sentence-based evaluation that used sentence likelihood difference (SLD) to calculate gender bias in LLMs\cite{Barza_2023}. Certification of fairness in AI systems was also considered by researchers who proposed a standard operating procedure (SOP) for fairness certification, Fairness Score and Bias Index, noting that different metrics would be needed to score pre-processing and in-processing and that the approach would be required to vary by use-case\cite{agarwal2023fairness}. Researchers found that specific algorithms scored better for one set of individual features than others, indicating a link between fairness evaluation and algorithm selection\cite{alam2019random}. 

\subsubsection{Trust \& Safety Evaluation}
The automated evaluation of trust and safety of AI systems was also considered by researchers\cite{Alhussainetal_2019}\cite{Khalilietal_2021}. Researchers proposed an automated trust scoring process that used machine learning to develop a trust value for their use case of file sharing in peer-to-peer networks, automating a process to score the technical safety and likelihood of the file being dangerous\cite{Alhussainetal_2019}. Additionally, researchers developed a process that combined privacy and fairness evaluation, scoring both and proposing a trade-off for accuracy for each\cite{Khalilietal_2021}.   

\subsection{Semi-automated Evaluation Methods}
This section covers approaches to scoring, which involve automated and manual steps. These methods are primarily in the area of fairness and compliance. They require a human at some stage, balancing automation and human efficiency. Researchers have shown the need to tailor evaluations by using case\cite{LeeandFloridi_2021}\cite{van2022algorithmic}\cite{Agarwal_2020} and to incorporate considerations such as cultural differences in fairness evaluation\cite{Paganoetal_2023}. In the case of healthcare, researchers reported that context was important in fairness evaluation for clinicians, noting a preference for a human-in-the-loop approach rather than a fully automated system\cite{ryan2023integrating}. 

\subsubsection{Fairness \& Compliance Evaluation}
Researchers have proposed several semi-automated evaluation methods for fairness and compliance in AI\cite{LeeandFloridi_2021}\cite{antunes2018fairness}\cite{sharma2020}\cite{nakao2023towards}\cite{stumpf2024need}. A number of these frameworks were automated methods of fairness evaluation combined with a human element to set thresholds or decide trade-offs between metrics. One approach included developing transparent processes that mapped trade-offs between metrics\cite{LeeandFloridi_2021}, while a second involved injecting controls, wrapping existing operations and extending workflow primitives\cite{antunes2018fairness}. A third method included allowing a human to define the fairness requirement, specifying assumptions and assertions so that the tester can generate inputs that satisfy these assumptions and violate assertions\cite{sharma2020}. A semi-automated user-centred approach to fairness evaluation called FairHIL (Fair Human-in-the-Loop) was developed that offers a visual user interface that provides a combination of visualisations including outcome features, feature intersection and causal graphs to help users identify bias and unfairness\cite{nakao2023towards}. Users can add labels and adjust the feature weighting to retrain the model until they achieve an acceptable user fairness outcome. The tool focuses on accessibility and explainability for non-AI experts. Researchers also evaluated the effects of cultural differences in users interacting with the FairHil tool\cite{stumpf2024need}. 

\subsubsection{Risk \& Accountability Evaluation}
One paper proposed a semi-automated method for risk evaluation. This structured method provides an open vocabulary for AI risks (VAIR)\cite{Golpayeganietal_2023}, facilitating the automation of AI risk category identification, a required step for AI assessment in the EU AI Act.

\section{Industry Tools for Evaluating TAI}
In addition to the aforementioned academic works in evaluating TAI, various industry tools are in use today that aim to ensure AI systems adhere to ethical, legal, and performance standards. The most commonly used tools are manual questionnaire-based tools such as the ALTAI\cite{EUALTAI_2020} and ISO/IEC 42001\cite{ISO42001}, which rely on self-assessments based on established principles, aligning with the self-assessment requirements of the EU AI Act\cite{eu2024aiact}. These tools rely on human judgment and expert evaluations to identify risks and compliance issues, ensuring a thorough, albeit time-consuming, evaluation process. These manual methods are often supplemented by frameworks such as the NIST AI Risk Management Framework\cite{nist_framework}, which provides comprehensive guidelines for assessing safety, fairness, and transparency. 

Automated assessment tools are becoming increasingly prevalent in the industry due to their efficiency and scalability. Tools like IBM's AI Fairness 360 and Microsoft Fairlearn are used to evaluate AI models for bias, fairness, and transparency without human intervention\cite{harris2023mitigating}. However, these are not accompanied by scientific, peer-reviewed papers evaluating their tools against the state-of-the-art works in this area\cite{johnson2023fairkit}. Johnson et al.\cite{johnson2023fairkit} publish an open-source toolkit called fair kit-learn, which is designed to support engineers in training fair machine learning models which found a better trade-off between fairness and accuracy than students using state-of-the-art tools sci-kit-learn and IBM AI Fairness 360\cite{johnson2023fairkit}.

These tools use sophisticated algorithms to identify and mitigate potential issues in AI systems, providing a scalable solution for large-scale AI deployments. Automated and semi-automated tools are particularly valuable, offering continuous monitoring and evaluation, enabling companies to maintain high standards of trustworthiness as AI systems evolve. Semi-automated tools such as Amazon SageMaker\cite{nigenda2022amazon} combine automated algorithms with human oversight, ensuring a balance between efficiency and expert insight. Amazon SageMaker continuously monitors real-time data, concepts, bias, and feature attribution drives in models. These tools require human intervention at critical stages to set parameters and make interpretive decisions, ensuring that ethical and fairness considerations are adequately addressed. 

Despite these advantages, recent research has highlighted several challenges practitioners face when using these tools. Practitioners find it difficult to translate real-world fairness concerns into quantifiable metrics that these toolkits can assess\cite{deng2022exploring}. There is also a need for toolkits to be able to integrate more seamlessly into existing ML pipelines and to provide more guidance and resources for responsible usage\cite{deng2022exploring}. Referring specifically to mitigating age bias in job selection using Microsoft Fairlearn and AI Fairness 360, researchers also found that significant human effort was required to make these toolkits work effectively to mitigate bias, making them impractical for usage in real-world applications\cite{harris2023mitigating}.

\section{Barriers to Trustworthy AI Evaluation}
The complexity required for a complete evaluation of TAI presents several challenges. The barriers to evaluating TAI found in the literature include the following: 

\paragraph{Diversity in Trustworthy AI Evaluation Method}
Evaluation methods exist for all aspects of TAI. However, the more mature areas of TAI have more advanced evaluation methods. For example, with several established methods, fully automated evaluation methods are available for fairness evaluation. Areas like risk and safety have some automatic and semi-automatic methods showing potential for more automation of technical aspects of AI where metrics are available. Evaluation approaches that considered less researched areas of TAI or holistic methods that considered multiple areas of TAI were primarily conception or manual methods.

\paragraph{Lack of Standardisation or Metrics for Evaluation}
Within the various TAI principles, there is a lack of consistency across all evaluation methods regarding what was being assessed. Even in similar industries using similar methods, the evaluation criteria or metrics used for evaluation were inconsistent. Regardless of the method used, this lack of consistency around evaluation criteria and metrics is a barrier to TAI evaluation and highlights a need to establish use case-specific benchmarks and acceptable thresholds for TAI evaluation. 

\paragraph{Use Case Specific Evaluation Methods Required}
Clinicians found that context was essential when deciding acceptable evaluations for AI fairness. AI systems are complex, and their design varies by use case. Due to this complexity, the evaluation method will vary by use case. For example, evaluating a decision-making AI system requires a different approach versus other AI use cases such as an LLMs. 

\paragraph{Human-in-the-Loop is Essential}
Although some automated methods exist to evaluate aspects of TAI, the semi-automated evaluation method is preferable if it integrates a human-in-the-loop. Additionally, due to a lack of maturity in many TAI principles, which have no metrics or automated methods for evaluation, a manual questionnaire-based stage is required for a comprehensive TAI evaluation. Additionally, even with more developed TAI principles such as fairness, a decision must be made manually to decide what is fair for the given use case. 

\paragraph{Discrepancies Between Stakeholders}
Researchers found that different stakeholders all required different levels of transparency, meaning different methods and criteria for evaluation may be required for various groups of stakeholders. There are additional discrepancies between what stakeholders, such as AI and law experts, consider fair and what a layperson considers fair. There have been some semi-automated approaches to establishing ethical norms that can include multiple perspectives to combat this. One proposed conceptual method\cite{buenfiletal_2019} involved extracting ethics from social media, which humans would then review for evaluation. Another approach was a semi-automated method\cite{nakao2023towards} involving the development of a user interface with TAI metrics agreed upon by the AI developer that enabled human stakeholders to evaluate, make adjustments and decide trade-offs between TAI metrics.

\paragraph{Auditing and Third Party Accreditation is Required}
The research showed a need for governance in TAI evaluations that involved some form of access to the AI system. Several researchers who published conceptual governance frameworks proposed the inclusion of a third-party accreditation body which did this. These bodies would aim to provide the needed audits and governance for TAI evaluation. The research showed the potential to automate the audit and certification process for some TAI principles based on agreed metrics and benchmarks.

\paragraph{Fragmented Development and Accountability}
AI systems built using multiple organisations, including third-party data providers, face significant evaluation barriers. AI producers may lack access to necessary information from contributing organisations which they require for comprehensive TAI evaluations. For example, AI trained on data purchased from a third party might lack insight into data consent and acquisition processes, hindering thorough evaluation. In such instances, the AI producer struggles to assume accountability for development steps outsourced to other entities, making it challenging to perform a complete TAI assessment.

\section{Future Directions for Trustworthy AI Evaluation}
To successfully evaluate TAI, the literature calls for future AI systems to have ongoing semi-automated evaluation capabilities. Successful prototypes include using transparent or explainable models, with an interface allowing human decision-making of thresholds, trade-offs and/or definitions to be input into the model. This can be done by an expert in the field or a third-party accreditation body. Universal evaluation criteria and thresholds do not apply from one use case to the next, meaning that TAI principles would need a specific evaluation criterion for each use case. 

There is a disconnect between the tools and research in this area. Tools used at the industry level have typically not been peer-reviewed and, when evaluated by researchers, are insufficient for comprehensive TAI evaluation versus the state of the art in the literature. 

The findings of this paper have significant implications for AI policy. The research underscores the necessity for standardised evaluation frameworks to assess the trustworthiness of AI systems. The current EU approach relies primarily on self-assessment and does not include methods or evaluation criteria for TAI evaluation, which the literature shows a clear need for. TAI standards developed by policymakers must be applied across use-case-specific AI applications to ensure ethical and fair practices. To facilitate comprehensive TAI evaluations for AI systems, governance frameworks in the literature propose third-party certification and standard methods and evaluation criteria, including metrics agreed upon by regulatory bodies based on their industry-specific needs and use cases. There is a disconnect between what policymakers, AI experts, and a standard non-expert user consider fair, along with differences based on culture, showing a need for more input from various laypeople to decide acceptable TAI evaluation approaches for individual use cases.

\paragraph{Authors and Affiliations}
All authors have reviewed and consented to the publication of the manuscript as presented. This research received partial support from the Science Foundation Ireland under grants 13/RC/2106P2 (ADAPT) and is co-funded by the European Regional Development Fund (ERDF).The data employed in this review are sourced from publicly available materials, including published research articles, ISO standards, books, and openly accessible databases and industry publications. All sources are duly cited and listed in the reference section of this paper. No new data was generated or collected specifically for this review.

\bibliography{Mybib}

\end{document}